\documentclass{article}

\usepackage{amsmath}
\usepackage{graphicx}
\usepackage{slashed}
\usepackage{bbold}
 \usepackage{relsize}
  \usepackage{verbatim}

 \usepackage{cite}
\usepackage{amssymb}
\usepackage{times,bbm}
\usepackage{graphics,dcolumn,bm,float}
\usepackage{amssymb,amsmath,rotate,color}
\usepackage{mathtools}
\usepackage{booktabs}
\usepackage{rotating}
\usepackage{float}
\usepackage{braket}
\usepackage{url} 
 \usepackage{hyperref}
 \usepackage{subcaption}
\usepackage{tikz}
\usepackage{pgfplots}

\newcommand{\bmt}{\left[\begin{matrix}}
\newcommand{\emt}{\end{matrix}\right]}

\newcommand{\ee}{\end{equation}}

\newcommand{\eea}{\end{eqnarray}}

\newcommand{\bigt}{\bigtriangleup}

\newcommand{\bfrp}[1]{{({\bf r}^\prime )}}

\usepackage{tikz}
\usetikzlibrary{arrows.meta, decorations.pathmorphing}

\begin{document}

\def\boxit#1{\vcenter{\hrule\hbox{\vrule\kern8pt
      \vbox{\kern8pt#1\kern8pt}\kern8pt\vrule}\hrule}}
\def\Boxed#1{\boxit{\hbox{$\displaystyle{#1}$}}} 
\def\sqr#1#2{{\vcenter{\vbox{\hrule height.#2pt
        \hbox{\vrule width.#2pt height#1pt \kern#1pt
          \vrule width.#2pt}
        \hrule height.#2pt}}}}
\def\square{\mathchoice\sqr34\sqr34\sqr{2.1}3\sqr{1.5}3}
\def\Square{\mathchoice\sqr67\sqr67\sqr{5.1}3\sqr{1.5}3}
\def\lambdabar{{\mathchar'26\mkern-9mu\lambda}}
\def\thrdotovervx{\buildrel\textstyle...\over v_x}
\def\thrdotovervy{\buildrel\textstyle...\over v_y}
\title{\bf 
Collective Modes in Weyl Superconductors and the Axial Anomaly
}

\author{{\small \ Mehran Z. Abyaneh}\footnote{mehran.z.abyaneh@gmail.com}
 \\
        {\small Department of Physics, Shahid Beheshti University, G.C.,
        Evin, Tehran 19839, Iran}}
\date{\today}
\maketitle
\begin{abstract}
We develop a covariant Lagrangian formulation for a time reversal symmetry broken, minimally relativistic three dimensional Weyl superconductor that preserves continuous chiral invariance in the chiral limit. Within this framework, the Fulde–Ferrell–Larkin–Ovchinnikov pairing spontaneously breaks the axial $U(1)_A$ symmetry, giving rise to distinct collective excitations. Using the Nambu–Jona–Lasinio approach, we identify a pseudo–scalar Nambu–Goldstone mode (This mode is absent for conventional BCS inter-node pairing) that acquires a small mass under explicit chiral–symmetry breaking and couples to gauge fields through the axial anomaly, closely analogous to the neutral–pion decay in QCD. Although this anomaly induced decay is strongly suppressed in the bulk by the Meissner effect, it may occur via surface electromagnetic fields. Our analysis also predicts additional vector and axial–vector collective modes associated with the broken $U(1)_A$ symmetry, providing a unified description of collective excitations in Weyl superconductors and their correspondence to mesonic modes in QCD.
\end{abstract}

\medskip
{\small \noindent PACS number:} 74.90.+n; 11.30.Rd; 11.15.Ex.\newline
{\small Keywords: Superconductivity; Weyl Materials; Chiral Symmetry Breaking; FFLO Pairing; Axial Anomaly; Spontaneous Symmetry Breaking}

\bigskip
\section{Introduction}
It is known that the $U(1)$ gauge symmetry is broken~\cite{Weinberg} via the spontaneous symmetry breaking (SSB) in the mean-field reduced Bardeen-Cooper-Schrieffer (BCS) Hamiltonian~\cite{BCS}. However, it can be restored via a phase mode called the Nambu-Goldstone (NG) mode~\cite{Nambu1, Nambu2}. The NG mode restores the broken $U(1)$ gauge symmetry via the radiative corrections to the vertex diagram~\cite{Littlewood} and can be absorbed into the longitudinal component of electromagnetic fields, getting elevated to the plasma frequency due to the Anderson-Higgs mechanism~\cite{Anderson1,Anderson2, Higgs} and leading to the Meissner effect~\cite{Meissner}.

Since in Weyl semimetals (WSMs) chirality can be treated as a conserved charge~\cite{Bednik2016} corresponding to a continuous $U(1)_A$ axial symmetry, the Bogoliubov–de Gennes (BdG) Hamiltonian of a three-dimensional Weyl superconductor (3DWS) respects $U(1) \times U(1)_A$ symmetry in the chiral limit. Hence,
  when both of the symmetries get spontaneously broken by an $s$-wave pairing, two NG modes should emerge: the familiar NG mode associated with the gauge $U(1)$ symmetry, and a new pseudo-scalar NG mode associated with $U(1)_A$ symmetry~\cite{AFM2025}. Also, the momentum-space separation of Weyl nodes in WSMs can be removed from the classical Lagrangian by a chiral (axial) gauge transformation. However, that transformation is not symmetry-allowed at the quantum level: the path-integral measure is not invariant. Its Jacobian produces the axial (chiral) anomaly.

Two types of pairings are possible in 3DWSs, one is the BCS inter-node pairing, in which electrons of opposite chirality form zero-momentum Cooper pairs; and the other is the FFLO intra-node pairing~\cite{FFLO}, where electrons with the same chirality form finite-momentum pairs. While the early modeling suggested FFLO pairing may be favorable in inversion-symmetric systems~\cite{Cho2012}, later studies~\cite{Bednik2015} showed that odd-parity BCS states are typically lower in energy unless inversion symmetry is explicitly broken. Although the BCS pairing has been realized in 3DWSs~\cite{Cho2012}, the FFLO pairing remains experimentally elusive~\cite{Matsuda2007}.

In the superconducting phase, the conventional BCS pairing mixes opposite-momentum quasiparticles and fully gaps the Weyl nodes, thereby suppressing the low-energy chiral anomaly and its associated transport signatures. In contrast, finite-momentum FFLO pairing can intertwine rather than eliminate the separated Weyl nodes, yielding Weyl excitations whose residual node separation encodes an effective axial gauge field, allowing modified anomaly-related responses to persist. Based on this peculiarity we observe that the pseudo-scalar NG mode can decay to two photons via the axial anomaly when the pairing is of FFLO type (see below).

On the other hand, the notion of SSB was transported to strong-interaction physics~\cite{NJL1,NJL2} by Nambu and Jona-Lasinio (NJL), motivated by the similarity between the Bogoliubov–Valatin equations~\cite{Valatin,Gennes,Bogoliubov1}, which define quasiparticle excitations in superconductors, and the Dirac equation~\cite{dirac2-1928}. In the NJL model, which is not derived from quantum chromodynamics (QCD) but addresses the SSB of the chiral symmetry, such an SSB takes place via the quark condensate $\langle \bar q q \rangle$, producing pions (pseudo-scalar mesons) as NG bosons~\cite{Scherer1,thesisAbyaneh}. Because quarks have small but nonzero bare masses, the pions acquire a finite mass, explained by the Gell-Mann–Oakes–Renner (GOR) relation~\cite{GOR}. Using the Fierz transformations, NJL model also predicts scalar, vector and axial-vector excitations in the particle spectrum. 

In the NJL model and in QCD, bound states come from the dynamics of interacting quarks like gluon-mediated interactions. Analogously, we expect collective modes to exist in Weyl semimetals if interactions drive the system into a symmetry-broken state i.e., via superconductivity, leading to bound states. Hence,
in this work, we study the collective modes in 3DWSs, employing an NJL-inspired framework. Despite differences between superconductors (where gauge symmetry is broken) and QCD (where it is preserved)~\cite{Weinberg} and also the nature of interactions, the analogy turns out to be fruitful. 

In this work, we write the BdG Hamiltonian in a covariant form ( in an 8-dimensional representation) which, leads to a new generator corresponding to the axial symmetry. This allows us to probe the ``fate" of the axial symmetry in the superconductive phase of 3DWSs, which to the best of our knowledge has remained unexplored in the literature.
We show that the SSB of $U(1)_A$ symmetry leads to a pseudo-scalar NG mode (which is analogous to the pion), a scalar amplitude mode\footnote{This amplitude mode (Higgs mode) is distinct from the one originating from the SSB of the $U(1)$ gauge symmetry~\cite{Littlewood}, which is directly observed~\cite{Shimino,Klein1984}.} plus a vector and an axial-vector excitation.
Furthermore, incorporating explicit chiral-symmetry-breaking perturbations in the effective Lagrangian (such perturbations can exist for example in TI–NI multilayer models of Weyl metals~\cite{Bednik2015}), we demonstrate that the pseudo-scalar NG mode becomes massive and can decay to two photons via the chiral anomaly. Since the SSB of the $U(1)_A$ symmetry is triggered by the FFLO pairing, such a decay can be a signature of the FFLO superconductivity. The results lead to a rich phenomenology, with clear experimental signatures that connect low-energy QCD analogies to tabletop condensed-matter realizations.

Several aspects of the present work build on earlier studies on axionic collective modes
\cite{Boyanovsky25,Shyta25,Mottola2024,Creti24},
NJL-type descriptions of interacting Weyl systems
\cite{Creti24,Boyanovsky25},
and anomaly-induced electromagnetic responses in Weyl superconductors
\cite{Bernabeu23,Lhachemi23,Palumbo25}.
In particular, the emergence of axion-like responses from axial symmetry
considerations and their relation to Chern--Simons terms has been discussed in
a variety of contexts. The novelty of the present work lies in, first a unified covariant BdG formulation incorporating
finite-momentum pairing, second the identification of spontaneous axial symmetry
breaking tied specifically to FFLO pairing in Weyl superconductors, and the last but not the least a systematic classification of collective modes within an $8\times8$ NJL framework.

The remainder of this paper is organized as follows: in Sec.~2, we briefly review spontaneous symmetry breaking in the NJL model. In Sec.~3, we write the BdG Hamiltonian for a time-reversal symmetry-broken 3DWS and construct the covariant Lagrangian. Sec.~4 provides a field-theoretic framework the emergence of collective modes, emphasizing the pseudo-scalar NG mode associated with FFLO pairing and its decay via the axial anomaly. We conclude in Sec.~5 with prospects for experimental observation.
\section{ NJL Model and Spontaneous Chiral Symmetry Breaking, a Brief reminder}\label{sec2}
\indent
For reference, we recall the essential ingredients of the Nambu--Jona-Lasinio (NJL) model. 
The Lagrangian for a single quark flavor reads~\cite{NJL1,NJL2} 
\begin{equation}
\mathcal{L}_{\rm NJL}
 = \bar{\psi}(i\gamma^{\mu}\partial_{\mu}-m_{0})\psi
 + g\big[(\bar{\psi}\psi)^2+(i\bar{\psi}\gamma^5\psi)^2\big],
\label{eq:NJLbasic}
\end{equation}
where $\hbar=1=c$ is assumed, $g$ is the coupling, $m_{_{0}}$ is the bare mass and the mass gap is going to be generated dynamically. 
Also, $\bar\psi=\psi^\dagger\gamma_0$ and $\gamma_{\mu}$ are
the Dirac gamma matrices. 

In the  limit $m_{_{0}}=0$, such a Lagrangian is invariant against the set of
two independent continuous global transformations, namely
\begin{eqnarray}\label{LNJL21}
\psi&\rightarrow& \exp{(i \alpha)}\psi,
\qquad\bar{\psi}\rightarrow\bar{\psi} \exp{(-i \alpha)},\\
\psi&\rightarrow& \exp{(i \gamma_5\alpha)}\psi,
\qquad\bar{\psi}\rightarrow\bar{\psi} \exp{(i\gamma_5
\alpha)}\, .
\end{eqnarray}
Due to the Noether's theorem there are two conserved currents
\begin{equation}\label{current0}
j_{\mu }  = \bar{\psi}\gamma_\mu  \psi;
\qquad
 j_{\mu }^5 =  \bar{\psi}\gamma_\mu\gamma_5 \psi;
 \end{equation}
 namely, the vector and axial vector currents, respectively, which satisfy the continuity equations
\begin{equation}\label{25}
\partial_{\mu}j_{\mu}=0\quad\textit{and}\quad\partial_{\mu}j_{\mu}^5=0\, \ .
\end{equation}
This corresponds to the conservation of the fermion number and the chiral or $\gamma_5$ charge, respectively. 

Spontaneous breaking of the axial symmetry generates a dynamical mass
\begin{equation}
m=-2g\langle\bar{\psi}\psi\rangle ,
\end{equation}
producing a massless NG mode i.e. the pion, together with 
a massive scalar mode (the $\sigma$ meson). 
Other Fierz channels yield vector and axial-vector collective modes, 
analogous to the $\rho$ and $a_{1}$ mesons. 
When a small bare mass $m_{0}$ is included, the NG boson becomes light but massive, following the Gell-Mann--Oakes--Renner (GOR) relation
\begin{equation}
m_{\pi}^{2} f_{\pi}^{2} = -m_{0}\langle\bar{\psi}\psi\rangle .
\label{eq:GOR}
\end{equation}

Although the NJL model is not derived from QCD, it captures essential features of spontaneous chiral symmetry breaking and collective excitations. 
The same mechanism will appear in our 3DWS model, where it serves as an analogous framework to describe the emergent bosonic modes in Weyl superconductors.
All intermediate steps and the standard derivations of Eqs.~(\ref{eq:NJLbasic})--(\ref{eq:GOR}) are summarized in Appendix~A. Also, a brief summary of the standard derivation of the pion mass and its anomaly-induced decay to two photons is provided in Appendix A for completeness.

\section{Covariant Form of the BdG Hamiltonian in Weyl Superconductors}\label{cov}
In this section we write the BdG Hamiltonian of a 3DWS in a covariant form. 
To proceed we consider a minimal relativistic low-energy model of a Weyl semimetal with two Weyl nodes of opposite chirality, separated by $ 2\,\mbox{\boldmath$q$} $ in momentum space. The time-reversal symmetry is broken by the chiral shift vector $ \mbox{\boldmath$q$} $, which encodes the momentum-space separation between the Weyl nodes. The Weyl Hamiltonian around the nodes located at $ \pm \mbox{\boldmath$q$} $ (in units where $ v_{_{\rm F}} = 1 $) takes the block-diagonal form
\begin{equation}\label{WeylH}
{\cal H}_{\rm W}(\mbox{\boldmath$p$} )=\left( \begin{array}{cc}
H^{\rm W}_{+} & \mathbf{0}_{2\times 2} \\
\mathbf{0}_{2\times 2} \  & H^{\rm W}_{-}\\
\end{array} \right)
\qquad\text{with}\qquad
H^{\rm W}_{\pm}= \pm
\mbox{\boldmath$\sigma$}\cdot\left(\mbox{\boldmath$p$}
\mp\mbox{\boldmath$q$}\right),
\end{equation}
where $\sigma_{i}$, with i=1,2,3, are the $2\times2$ Pauli matrices in the spin space,  $ \mbox{\boldmath$p$} $ is the quasiparticle momentum, the signs $ \pm $ refer to the two chiralities and $\mathbf{0}_{2\times 2}$ is a 2 dimensional zero matrix.
Although the momentum-shift term $\mathbf{q}$ can be gauged away from the kinetic term, coupling to external electromagnetic fields makes this chiral rotation non-symmetry-preserving, generating axion-like contributions to the effective action (see below), including anomalous Hall and chiral magnetic effects~\cite{ZyuzinBurkov2012,Burkov2}.
 Then, introducing $4\times4$ Dirac matrices in the chiral (Weyl) basis as $\gamma_0 = \tau_1 \otimes \mathbb{I}_2$,
$ \gamma_i = i\,\tau_2 \otimes \sigma_i$ with $(i=1,2,3)$ and $\gamma^5=i\gamma^0\gamma^1\gamma^2\gamma^3$, where the set of Pauli matrices $\tau_i$ act in chirality space,
the Weyl Hamiltonian reduces to
\begin{equation}\label{WeylHD}
{\cal H}_{\rm W}(\mathbf p)
=- \gamma_0\,\boldsymbol\gamma \cdot (\mathbf p-\mathbf{q} \gamma_5).
\end{equation}

Superconductivity can emerge intrinsically or via proximity effect in WSMs such as MoTe$_2$~\cite{Liu2016,Jiang2017} or TaP~\cite{Xu2015}. The BdG Hamiltonian can be written, including both chiralities, in the form
\begin{equation}\label{BdGH}
\mathcal{H}_{\rm BdG}(\mathbf{p}) =
\begin{pmatrix}
\mathcal{H}_{\rm W}(\mathbf{p}) - \boldsymbol{\mu} & \hat{\Delta}\\
\hat{\Delta}^\dagger & \boldsymbol{\mu} - \mathcal{C}^{-1} \mathcal{H}_{\rm W}(\mathbf{p}) \mathcal{C}
\end{pmatrix}, 
\end{equation}
where $\mu$ is the chemical potential and $\mathcal{C}=i \gamma_0\gamma_2 K$ is the charge-conjugation operator with K the complex conjugation. The $4\times 4$ pairing matrix $\hat{\Delta}$ couples the positive- and negative-energy sectors and is generated dynamically through spontaneous symmetry breaking. We continue with $\mu=0$ as it is common to choose for undoped WSMs~\cite{Dutta2020}. 
Although the two pairing sectors have been dealt with
separately in the literature, see, e.g., Ref.~\cite{Sinha2020}, including both chiralities in one representation, the BdG Hamiltonian~(\ref{BdGH}) writes as
\begin{equation}
\mathcal{H}_{_{\rm BdG}}=\left(
\begin{array}{cccc}
- \gamma_0\,\boldsymbol\gamma \cdot (\mathbf p-\mathbf{q} \gamma_5)& \hat{ \Delta}\\ 
 \hat{ \Delta}^{\dagger}&- \gamma_0\,\boldsymbol\gamma \cdot (\mathbf p+\mathbf{q} \gamma_5)\\
\end{array}\right),
\label{BCS/FFLO}
\end{equation}
and acts on the Nambu basis
\begin{equation}\label{nambuspinor}
\Psi =(\Phi_-,\Phi_+,\Phi_-^c,\Phi_+^c)^T,
\quad\textit{with}\quad\Phi_\pm=(\phi_{\uparrow \pm}, \phi_{\downarrow \pm})^T, \quad \Phi_\pm^c =(\phi^*_{\downarrow \pm},-\phi^*_{\uparrow \pm})^T.
\end{equation}
 Depending on weather the pairing amplitude is of FFLO or BCS type we can write
 \begin{equation}\label{pairings}
  \hat{\Delta}_{_{\rm F}}\! \equiv\!  \left(
\begin{array}{cccc}
\Delta_{_{\rm F}}&\mathbf{0}_{2\times 2} \\\
\mathbf{0}_{2\times 2} & \Delta_{_{\rm F}} \\
\end{array}
\right)\quad\textit{and}\quad
  \hat{\Delta}_{_{\rm B}}\! \equiv\!  \left(
\begin{array}{cccc}
\mathbf{0}_{2\times 2}&\Delta_{_{\rm B}} \\\
\Delta_{_{\rm B}} &\mathbf{0}_{2\times 2}\\
\end{array}
\right),
  \end{equation}
   where ${ \Delta}_{_{\rm F}}$ and ${ \Delta}_{_{\rm B}}$ are two dimensional diagonal matrices with the indices F and B standing for the FFLO and BCS pairings, respectively. 
   The BCS pairing corresponds to pairing between opposite
chiralities and therefore is momentum-independent, whereas the FFLO pairing is intra-node and naturally carries finite center-of-mass momentum.
Equation~(\ref{BCS/FFLO}) is written in a gauge where the finite-momentum
character of the FFLO state is encoded in the axial shift
$\mathbf p\to\mathbf p\pm\mathbf q\gamma_5$, such that the FFLO pairing amplitude
$\hat\Delta_{\rm F}$ is spatially uniform. 

  Now, defining the 8 dimensional matrices
\begin{align}\label{8Gamma}
\Gamma_\mu=\rho_1\otimes\gamma_\mu, \quad \Gamma_5 = \mathbb{I}_{{2}}\otimes\gamma_5,
\quad\textit{and}\quad
\Sigma_i=\rho_i\otimes \mathbb{I}_{{4}},
 \end{align}
 where $\rho_i$ with i=1,2,3 are Pauli matrices in the particle-hole basis and  $\mathbb{I}_{{4}}$ is a $4\times 4$ identity matrix, the BdG Hamiltonian, and the corresponding Lagrangian in the FFLO and BCS channles\footnote{The enlarged $8\times8$ Dirac--Nambu structure provides a convenient
symmetry-based and algebraic framework for treating different pairing channels
on equal footing, but does not imply their simultaneous physical realization.}, can be written in the covariant form 
\begin{align}\label{covHamil}
\mathcal{H}_{{\rm F}}
= {\mathbf A}\cdot(\mathbf p+\mathbf q\Gamma_5)+\bigt_{\rm F}\Sigma_1,
\qquad
\mathcal{L}_{\rm F}
=\overline{\Psi}\big(\Gamma_\mu(p^\mu+q^\mu\Gamma_5)-\bigt_{\rm F}\Omega\big)\Psi .\end{align}
and
\begin{align}\label{covHamilBCS}
\mathcal{H}_{{\rm B}}
= {\mathbf A}\cdot(\mathbf p+\mathbf q\Gamma_5)+\bigt_{\rm B}\Gamma_0,
\qquad
\mathcal{L}_{\rm B}
=\overline{\Psi}\big(\Gamma_\mu(p^\mu+q^\mu\Gamma_5)-\bigt_{\rm B}  \mathbb{I}_{{8}} \big)\Psi,
\end{align}
 where $\bigt_{_{\rm F}}$ and $\bigt_{_{\rm B}}$ are scalar $s$-wave pairings. The operator $A_i \equiv -\Gamma_0 \Gamma_i$ ensures the correct particle-hole and spin structure and leads to the kinetic part of the BdG Hamiltonian in the $8\times 8$ Nambu basis, $\overline{\Psi}=\Psi^\dagger \Gamma_0$ and $\Omega\equiv\Gamma_{0}\Sigma_1$. 
 Equations~(\ref{covHamil}) and~(\ref{covHamilBCS}) do not introduce any additional approximation or
gauge choice, but merely rewrites the BdG Hamiltonian
(\ref{BCS/FFLO}) in a compact covariant form.
The axial momentum shift proportional to $\mathbf q$, is encoded in the
$8\times8$ representation by the matrix $\Gamma_5= \mathbb{I}_{{2}}\otimes\gamma_5$.
In this way, the finite-momentum character of the FFLO state is entirely captured
by the $\Gamma_5$-dependent kinetic term, while the pairing amplitude
$\bigt_{\rm F}$ remains uniform. Also, the derivation of the covariant form and the Lorentz invariance of Hamiltonian and Lagrangians in~(\ref{covHamil}) and~(\ref{covHamilBCS}) is discussed in the Appendix B.
\footnote{We stress that although the BdG Hamiltonian and Lagrangian can be written in a formally Lorentz-covariant 8$\times$8 Dirac form, 
 this covariance is algebraic that is, the actual low-energy system resides on a lattice with fixed Weyl node separation $\mathbf{q}$,
and true Lorentz symmetry is broken. The covariant structure provides a convenient framework to classify collective modes and
to identify anomaly-related couplings, but should not be interpreted as a physical spacetime symmetry.}

The 8$\times$8 Dirac representation of the BdG Hamiltonian, in the normal state, is formally invariant against two $U(1)$ transformations, generated by $\Sigma_3$ (corresponding to the gauge symmetry) and $\Gamma_5$, which corresponds to independent phase rotations of the left- and right-handed Weyl fermions (chiral symmetry).  
However, in the superconducting FFLO state, the pairing term $\bigt_{\rm F}\Sigma_1$ spontaneously breaks both $U(1)$ symmetries, because it couples left- and right-handed components as well as the particle-hole sectors. 
The breaking of the chiral symmetry gives rise to the pseudo-scalar mode associated with oscillations of the relative phase of the pairing condensate. On the other hand, the the BCS pairing term only breaks the $U(1)$ gauge symmetry (see Appendix B), meaning that such a pseudo-scalar mode only arises for the FFLO type of pairing and is absent for conventional BCS inter-node pairing due to its symmetry properties.

\section{The Nambu–Jona-Lasinio Model in Weyl Superconductors}
As discussed in sec.~(\ref{sec2}), the SSB of $U(1)_A$ in the NJL model gives rise to pions as the pseudo-scalar NG modes to restore the broken symmetry. Since the chiral symmetry is also broken spontaneously by $\bigt_{\rm F}$ (see Appendix B) in the context of 3DWSs, it is natural to construct an NJL-type Lagrangian with four-fermion interactions for such a system. Hence, to systematically analyze collective excitations in Weyl superconductors, we adopt a NJL-type formulation in the FFLO pairing channle.
This construction reproduces the phenomenological FFLO pairing structure, while providing a symmetry-based and algebraically systematic framework for classifying low-energy excitations. Focusing only on chiral symmetry for the sake of illustration, we can build a Lagrangian which is invariant under the chiral transformation~(\ref{transform11})
\begin{equation}\label{BdGNJLLAG}
{\cal L}_{\rm NJL}^{\rm 3DWS} = \bar{\Psi} (i \Gamma_\mu \partial^\mu -  M_0  ) \Psi 
+ \Omega \Big[G_S (\bar{\Psi} \Psi)^2 - G_P (\bar{\Psi} \Gamma_5 \Psi)^2\Big],
\end{equation}
where $M_0$ is an explicit symmetry-breaking mass term, and $G_S = G_P = G$ are the couplings in the scalar and pseudo-scalar channels, respectively.
Applying the mean-field approximation, when $M_0=0$, we get
\begin{align}
\Omega\,G \,(\bar{\Psi} \Psi)^2 \to \bar{\Psi} \Omega\bigt_{_{\rm F}}\Psi,
\end{align}
where the mass $- 2 G \langle \bar{\Psi}\Psi \rangle$ is generated dynamically using which, Lagrangian~(\ref{BdGNJLLAG}) reduces to Lagrangian in~(\ref{covHamil}).
Applying Fierz transformations (As discussed in Appendix C) and analyzing the T-matrix 
in
 \begin{equation}
\Gamma_S = \mathbb{I}_{{8}}, \quad \Gamma_P =  \Gamma_5, \quad \Gamma_V = \Gamma_\mu  , \quad \Gamma_A = \Gamma_\mu  \Gamma_5,
\end{equation}
channels, we can find a scalar phase mode and an amplitude mode in the scalar channel $\Gamma_S$ (well-known in conventional superconductors~\cite{Littlewood}), a pseudo-scalar phase mode (NG mode) and an amplitude mode in $\Gamma_P$ channel,  a vector mode in $\Gamma_V$ and an axial-vector mode in $\Gamma_A$ channel. The choice of interaction channels in the NJL model is dictated by the underlying symmetries and by the structure of the $8\times8$ Dirac representation, rather than being a unique microscopic prescription. 
Within this framework, the pseudo-scalar mode and other collective excitations naturally emerge as fluctuations of the pairing order parameters in the corresponding 
symmetry channels. 

For instance, we can write the polarization function for the pseudo-scalar channel, using the quasi-particle propagator $S(P)=(P^{\mu}\Gamma_{\mu}+  \bigt_{_{\rm F}} \Omega)/(P^2- \bigt_{_{\rm F}}^2+i \epsilon)$, as
\begin{equation}
\Pi^{\rm 3DWS}(Q) =- i \int \frac{d^4 P}{(2\pi)^4} \mathrm{Tr} \left[  \Gamma_5 \frac{i (\Gamma_\mu (P+Q)^\mu +  \bigt_{_{\rm F}}\Omega)}{(P+Q)^2 - \bigt_{_{\rm F}} ^2 + i \epsilon}  \Gamma_5 \frac{i (\Gamma_\nu P^\nu + \bigt_{_{\rm F}} \Omega)}{P^2 - \bigt_{_{\rm F}} ^2 + i \epsilon} \right].
\end{equation}
Using commutation relations~(\ref{commut})
and evaluating the trace at $Q^2 = 0$ gives
\begin{equation}\label{PiWS}
\Pi^{\rm 3DWS}(0) = 8 i \int^\Lambda \frac{d^4 P}{(2\pi)^4} \frac{1}{P^2 - \bigt_{_{\rm F}}^2 + i \epsilon}.
\end{equation}
Combined with
 the gap equation 
\begin{equation}
\bigt_{_{\rm F}} =-2 G Tr \Big[  \Omega \langle \bar{\Psi} \Psi\rangle\Big]=16 i G \bigt_{_{\rm F}}  \int^\Lambda \frac{d^4 P}{(2\pi)^4} \frac{1}{P^2 - \bigt_{_{\rm F}} ^2 + i \epsilon},
\end{equation}
we get
\begin{equation}\label{gapeq}
1 = 16 i G  \int^\Lambda \frac{d^4 P}{(2\pi)^4} \frac{1}{P^2 - \bigt_{_{\rm F}}^2 + i \epsilon},
\end{equation}
which implies
\begin{equation}
2 G \, \Pi^{\rm 3DWS}(0) = 1,
\end{equation}
signaling the presence of a {\em massless} pseudo-scalar NG mode~\cite{AFM2025}, with $K=2G$.
Deriving the masses of the other collective modes is beyond the scope of this work but can be done in a similar fashion. As in the NJL model, vector and axial–vector modes are expected to be heavier than scalar and pseudo-scalar modes, with masses of order the pairing scale or higher. This can be probed experimentally via spectroscopies sensitive to frequency and momentum, such as Raman scattering, THz spectroscopy, neutron scattering, or pump-probe techniques. 

Although our analysis makes use of a covariant $8\times8$ Dirac and NJL-type
formalism, the pseudo-scalar collective mode discussed here has a direct
interpretation within the Bogoliubov--de Gennes description of a Weyl
superconductor.
Physically, the pseudo-scalar mode corresponds to a relative phase fluctuation
between chiral pairing components in the finite-momentum superconducting state,
rather than a fundamental axion field, and its observability depends on material
parameters, disorder, and electromagnetic screening.

It is also distinct from the usual scalar Nambu–Goldstone phase mode and
represents an internal collective excitation of the superconducting order
parameter. We emphasize that the existence of this relative-phase mode, and the other collective modes discussed above, does not rely on
the formal extension to an $8\times8$ Dirac algebra.
The extended representation provides a compact and symmetry-transparent way
to classify collective excitations, but the underlying modes can be identified
directly at the level of the BdG Hamiltonian as a fluctuation in the internal
structure of the pairing order parameter.

\subsection{Explicit Chiral Symmetry Breaking}
In sec.~(\ref{cov}) we demonstrated that when the pairing potential is of the FFLO type, both the chiral and gauge symmetries get spontaneously broken. In the previous section we showed that the SSB of the chiral symmetry leads to a massless pseudo-scalar mode, which is analogous to pions in the NJL model.
We now examine the effect of a non-vanishing explicit chiral symmetry breaking term,  $M_0$ in Lagrangian~(\ref{BdGNJLLAG}), when only the FFLO pairing is present. Such a bare mass term breaks both rotations $\Psi\to\!\!e^{\, i \Gamma_5 \theta/2} \Psi$ and $\Psi \to e^{\, i \Sigma_3 \varphi/2} \Psi$. Focusing only on the chiral symmetry and expanding the chiral rotation for small $\theta$ we get
\begin{equation}
\delta {\cal L}_{\rm NJL}^{\rm 3DWS} 
= -M_0 \, \bar{\Psi} 
\Big( 1 + i \Gamma_5 \theta - \tfrac{1}{2}\theta^2 + \cdots \Big) \Psi .
\end{equation}
Comparing with relation~(\ref{chiralrot1}), and following the same procedure that leads to~(\ref{GOR}), we obtain the superconducting analogue of the GOR relation
\begin{equation}\label{spionmass}
M_{\rm ps}^2 \, F_{\rm ps}^2 = M_0\, \langle \bar{\Psi}  \Psi \rangle,
\end{equation}
 where the index ${\rm ps}$ refers to the pseudo-scalar collective mode.

Motivated by the formal analogy with NJL models in relativistic field theory,
we obtain order-of-magnitude estimates for the mass and decay rate of this
mode.
We stress, however, that this analogy is used only as a symmetry-based guide,
and not as a claim of dynamical equivalence with QCD or pion physics.
To estimate $M_{\rm ps}$, we recall that the binding energy of Cooper pairs in a FFLO superconductor is of order $\bigt_{\rm F}$, the superconducting gap. By analogy with QCD case discussed in Appendix A, up to dimensionless coefficients of order unity we identify $F_{\rm ps}\approx\bigt_{\rm F}$ and $\langle \bar{\Psi} \Psi \rangle \approx\bigt_{\rm F}^3$. Taking $M_0 \sim 10^{-3}\bigt_{\rm F}$ (consistent with the ratio of bare quark mass to $\Lambda_\chi$ in Appendix A) and adopting the estimate $\bigt_{\rm F} \approx 1~\text{meV}$~\cite{Ding}, relation~(\ref{spionmass}) gives
\begin{equation}\label{pionmass3}
M_{\rm ps} \;\approx\; 3.3 \times 10^{-2} \; \text{meV}.
\end{equation}
Thus, the pseudo-scalar mode acquires a small but finite mass under explicit chiral symmetry breaking, directly analogous to the pion in QCD. 

From a physical standpoint, the pseudo-scalar mode should be regarded as an
internal collective excitation of the superconducting condensate.
Its experimental visibility depends on its coupling to electromagnetic
fields, quasiparticles, and disorder.
In particular, the anomaly-induced coupling discussed below provides a
mechanism by which this mode may couple to external probes, although a
detailed quantitative prediction requires a material-specific microscopic
analysis beyond the scope of the present work.

\subsection{Axial anomaly and induced topological term}
\label{subsec:axialanomaly}

In QCD, a chiral rotation that removes a complex quark mass term generates the topological $\theta$--term~\cite{Scherer1}
\begin{equation}\label{tetaterm}
\mathcal{L}_{\theta} = 
\frac{\theta e^{2}}{32\pi^{2}} 
\, \epsilon^{\mu\nu\rho\sigma} F_{\mu\nu} F_{\rho\sigma},
\end{equation}
where $\theta$ is the axion angle, $e$ the electric charge, 
$F_{\mu\nu} = \partial_{\mu} A_{\nu} - \partial_{\nu} A_{\mu}$ is the electromagnetic field-strength tensor
and $\epsilon^{\mu\nu\rho\sigma}$ is the fully antisymmetric Levi--Civita tensor in four dimensions. This term encodes the nontrivial $\theta$--vacuum structure of the gauge field and represents the axial anomaly in QCD.

As shown in Refs.~\cite{Ishikawa1984,Goryo1999}, an analogous mechanism occurs in time--reversal--breaking 3DWSs, which can be viewed as minimally relativistic $(3\!+\!1)$--dimensional systems with four--component Nambu spinors. The Weyl--node separation $q_{\mu}=(0,\mathbf{q})$ can be gauged away via a local chiral rotation, in the absence of electromagnetic fields.  
When coupling to gauge fields is included, however, the fermionic path--integral measure is not invariant under this transformation.  
The Jacobian yields an additional contribution to the effective action,
\begin{equation}\label{eq:CSterm}
S_{\mathrm{CS}} = 
\frac{e^{2}}{4\pi^{2}}\!\int\!d^{4}x\, 
\epsilon^{\mu\nu\rho\sigma}\, q_{\mu} A_{\nu}\partial_{\rho}A_{\sigma},
\end{equation}
which is the three--dimensional Chern--Simons term responsible for the anomalous Hall and chiral magnetic effects in Weyl superconductors~\cite{Ishikawa1984,Goryo1999,ZyuzinBurkov2012,Burkov2}.  

Although the BdG Hamiltonian is written in a gauge where the FFLO order parameter
is uniform and the finite-momentum character is encoded in the axial shift
$q_\mu$, collective phase fluctuations of the FFLO condensate are introduced by
promoting $\bigt_{\rm F}\to\bigt_{\rm F}e^{i\theta(\mathbf r)}$.
A local chiral rotation of the Nambu spinor,
$\Psi \rightarrow e^{-i\theta(\mathbf r)\Gamma_{5}/2}\Psi$,
removes the phase of the pairing term at the expense of introducing an
additional axial gauge field $\tfrac{1}{2}\partial_\mu\theta$ in the fermionic
kinetic term, while the Weyl--node separation $q_\mu$ remains a fixed background
axial field.

From the viewpoint of effective field theory, the connection between the chiral rotation in Eq.~(\ref{tetaterm}) and the induced Chern--Simons term~(\ref{eq:CSterm}) can be compactly expressed by a Wess--Zumino--Witten (WZW) action that captures the anomaly inflow between four and three dimensions.  
In this picture, the gradient of the axion angle $\partial_{\mu}\theta$
enters the effective action in the same way as an axial gauge field and therefore
plays the same functional role as the Weyl--node separation $q_{\mu}$ (up to conventional normalization) in
3DWSs, although the two have distinct microscopic origins.
This correspondence underlies the topological origin of the electromagnetic
response and justifies the NJL--type analogy used in the present work.

\subsection{Lifetime Estimation}
As discussed above breaking of $U(1)_A$ symmetry in 3DWSs via SSB leads to a pseudo-scalar NG mode, which also becomes massive due to explicit chiral symmetry breaking.
Consequently, analogous to pions (as discussed in Appendix A ), such a pseudo--NG mode is expected to couple anomalously to the electromagnetic
field and, at the level of effective field theory, admits a radiative decay
channel into two photons via the axial anomaly.
The anomaly-driven radiative decay of the pseudo-scalar collective mode has the same operator structure as $\pi^0\to 2\gamma$ in QCD and hence, as an effective-field-theory estimate
of the anomaly-induced coupling strength, rather than a direct prediction of a
bulk decay process in a fully gapped superconductor, we can write its decay width in an analogous form with~(\ref {decayrate1}) as
\begin{align}\label{decrate}
\Gamma^{\rm D}_{\rm ps}
\simeq C\;\frac{\alpha_{\rm WS}^{2}\,M_{\rm ps}^{3}}{576\pi^{3}F_{\rm ps}^{2}}.
\end{align}
Here, $\alpha_{\rm WS}=(e^\ast)^2/(4\pi\hbar v_F)$, when units are recovered, is the effective dimensionless electromagnetic coupling of the quasiparticles (with $v_F$ their characteristic velocity and $e^\ast$ their effective charge), $F_{\rm ps}$ is the pseudo-scalar stiffness (playing the role of a decay constant), and $C$ is a dimensionless material dependent prefactor.  

For an order-of-magnitude estimate, we use the anomaly-driven form relation~(\ref{decrate}) with $C \approx 1$, $M_{\rm ps}\approx 3.3\times10^{-2}\ {\rm meV}$, $F_{\rm ps}\approx1\ {\rm meV}$, $v_F=10^{6}\ {\rm m/s}$, and $\alpha_{\rm WS}\!\approx\!0.22$ to get
 \begin{align}\label{decrate1}
\Gamma^{\rm D}_{\rm ps}\!\approx\!9.74\times10^{-14}\ {\rm eV},
\end{align}
corresponding to a decay rate of 
$1.48\times10^{2}\ {\rm s^{-1}}$ 
or a lifetime 
$\tau\!\approx\!6.8\times10^{-3}\ {\rm s}$. Because photon propagation in a superconductor is screened by the Meissner
effect~\cite{Meissner} or, equivalently, by the Anderson--Higgs mechanism
\cite{Anderson1,Anderson2,Higgs}, this decay channel is strongly suppressed in
the bulk. The estimate above therefore characterizes the intrinsic
anomaly-induced coupling of the pseudo-scalar mode rather than an observable
bulk lifetime. Nevertheless, the anomaly may still manifest near surfaces or
interfaces, where Higgsing is incomplete, leading to nonlinear optical responses
\cite{Morimoto2016}.

 In summary, while the existence and symmetry properties of the pseudo-scalar
mode are robust consequences of the pairing structure and symmetry breaking,
we emphasize that the anomaly-induced coupling discussed here should be viewed
as an effective low-energy description dictated by symmetry and topology.
While the operator structure of the coupling is fixed by the axial anomaly,
quantitative estimates of the pseudo-scalar mass and decay rate are necessarily
model-dependent and should be regarded as order-of-magnitude estimates rather than precise predictions.

\section{Discussion and Conclusion}

We have examined the consequences of spontaneous breaking of the axial $U(1)_A$ symmetry in a time–reversal–symmetry broken, minimally relativistic three–dimensional Weyl superconductor realized via FFLO pairing.
Starting from the BdG Lagrangian written in a chiral–invariant form; closely analogous to the NJL framework for low–energy QCD; we have analyzed the $T$–matrix in various interaction channels and found a rich collective–mode spectrum, including a pseudo–scalar NG mode. Under explicit chiral–symmetry breaking, this NG mode acquires a small mass and couples to gauge fields through the axial anomaly, in direct analogy with the neutral–pion decay in QCD. Although its decay, with an estimated lifetime of $\tau\!\sim\!10^{-2}$ s, is strongly suppressed in the bulk by the Meissner effect, it may proceed via surface photons or emergent electromagnetic fields. Such an anomaly–induced decay could serve as a characteristic experimental signature of FFLO pairing, whose unambiguous observation remains outstanding.

Beyond the pseudo–scalar channel, our analysis predicts additional collective excitations associated with the $U(1)_A$ symmetry breaking, including vector and axial–vector modes analogous to mesonic resonances in QCD. While the scalar and pseudo–scalar channels dominate the low–energy dynamics, the vector and axial–vector modes may appear in Bogoliubov–quasiparticle recombination spectra, offering further experimental avenues.
We explicitly highlight the following novel aspects of this work:
\begin{enumerate}
    \item A unified covariant BdG formulation for time-reversal-symmetry-broken Weyl superconductors including both the FFLO and BCS pairing channels.
    \item Identification of spontaneous axial $U(1)_A$ symmetry breaking associated with FFLO pairing.
    \item Systematic classification of collective modes using the $8\times8$ NJL structure, including the pseudo-scalar mode.
    \item A symmetry-based argument for the anomaly-induced coupling of the pseudo-scalar mode to electromagnetic fields.
\end{enumerate}
In summary, the present work establishes a direct correspondence between topological superconductivity, axial anomalies, and low–energy QCD. This analogy not only demonstrates the universality of spontaneous symmetry breaking and anomaly phenomena across disciplines but also suggests concrete experimental probes of neutral collective excitations in Weyl superconductors.
The purpose of this work is not to claim a direct dynamical equivalence between
Weyl superconductors and relativistic quantum field theories, but rather to
demonstrate how symmetry-based and anomaly-related concepts can be imported in
a controlled way to organize collective excitations in finite-momentum paired
superconductors.

\section*{Appendix A: Pion mass and anomaly-induced decay in the NJL model}
First, we write the NJL Lagrangian~\cite{NJL1,NJL2} as
\begin{eqnarray}\label{LNJL1}
{\cal L}_{\rm
NJL}=\bar\psi( i \gamma^{\mu}\partial_{\mu}-m_{_{0}})\psi+\sum_{\rm a}[g_{_{\rm S}}(\bar\psi\tau_{\rm a}\psi)^2+g_{_{\rm P}}(i\bar\psi
\gamma^5\tau_{\rm a}\psi)^2]\, ,
\end{eqnarray}
 where $\hbar=1=c$ is assumed, $\tau$ are the Pauli matrices in the isospin space, $g_{_{\rm S}}=g_{_{\rm P}}=g$ is the coupling in the indices S and P stand for scalar and pseudo-scalar channels, $m_{_{0}}$ is the bare mass and the mass gap is going to be generated dynamically. 
Also, $\bar\psi=\psi^\dagger\gamma_0$ and $\gamma_{\mu}$ are
the Dirac gamma matrices and  in the Weyl or chiral representation. 
Also the $\gamma_5$ matrix is constructed as $\gamma^5=i\gamma^0\gamma^1\gamma^2\gamma^3$
and the $\gamma_{\mu}$ matrices
satisfy
\begin{equation}\{\gamma_{\mu},\gamma_{\nu}\}\equiv \gamma_{\mu} \gamma_{\nu}+\gamma_{\nu} \gamma_{\mu}=2\,\theta_{\mu\nu} \mathbb{I}_{4},
\label{Eq:Gamma_1}
\end{equation}
where $\theta_{\mu\nu}$ (with $\mu,\nu= 0,\cdots,3$) is the
Minkowski metric in $(1+3)$ dimensions with the signature $-2$ and $ \mathbb{I}_{4}$ is the $4\times4$ unit matrix.

One can expand the Lagrangian~(\ref{LNJL1}) using the Fierz transformation, in the single quark flavor space, to find~\cite{NJL1,NJL2}
 \begin{equation}\label{LNJL2}
 {\cal L}'_{\rm
NJL}=-g_{_{\rm V}}(\bar\psi\gamma_{\mu}\psi)^2+g_{_{\rm A}}(\bar\psi
\gamma_{\mu}\gamma^5\psi)^2,
  \end{equation}
where $g_{_{\rm V}}=g_{_{\rm A}}=g'$ are the coupling in the vector and axial-vector channels, and these new terms read to vector and axial-vector interactions. Indeed, one can construct the meson T-matrix by solving the Bethe-Salpeter equation at a given squared four-momentum $ q^2$ of the quark-anti-quark pair to get~\cite{Vogl}
 \begin{equation}
 T(q^2)=k+k \Pi(q^{2}) k+...=\left[1-k  \Pi(q^{2})\right]^{-1} k\, ,
 \end{equation}
 where $k$ is the Born term of the T-matrix.
Here, $ \Pi$ represents the quark loop polarization function
\begin{equation}\label{polar1}
 \Pi_Y(q^{2})=i\int^{\Lambda} \frac{d^4p}{(2\pi)^4} Tr[\gamma_{_{Y}} s(p+q)\gamma^{\dagger}_{_{Y}} s(p)],
\end{equation}
with $s(p)=(p^{\mu}\gamma_{\mu}+m)/(p^2-m^2+i \epsilon)$ the quark propagator, $\Lambda$ the ultraviolet cutoff and the index Y standing for the specific channel, namely
\begin{align}
&\gamma_{_{\rm S}}=1,\qquad\gamma_{_{\rm P}}=\gamma_5,\qquad\gamma_{_{\rm V}}=\gamma_{\mu}\qquad\textit{and}\qquad
\gamma_{_{\rm A}}=\gamma_{\mu}\gamma_5.
\end{align}
 The meson mass $m_{_{\rm M}}$ is then found by solving the pole equation
\begin{equation}\label{pole}
1-k\Pi(q^2=m_{_{\rm M}}^2)=0,
\end{equation}
 in a given channel.
For example, the pseudo-scalar polarization function reads
\begin{equation}
\Pi(q^2=0)=4i\int^{\Lambda} \frac{dp^4}{(2\pi)^{4}}\frac{1}{p^{2}-m^{2}+i \epsilon}.
\end{equation}
On the other hand, the gap equation of the NJL model, $m= -2g\langle \bar{\psi}\psi\rangle=2g i {\rm Tr} [s(0)]$, yields
\begin{equation}\label{gapeq}
1=8 gi\int^{\Lambda} \frac{dp^4}{(2\pi)^{4}}\frac{1}{p^{2}-m^{2}+i \epsilon},
\end{equation}
via which 
\begin{eqnarray}\label{poleeq}
2g\, \Pi(q^2=0)=1,
\end{eqnarray}
which indicates the presence of a {\em massless} pseudo-scalar boson with $k=2g$. The procedure just explained can be followed again for the scalar, vector and axial-vector parts of the interaction Lagrangian to find other collective
states with masses proportional to the mass of their building blocks, i.e. the quarks~\cite{NJL1,NJL2}. By studying the collective excitations of the NJL model, one finds that
different interaction channels give rise to distinct bound states with
characteristic spin and parity. 

\subsection*{The Explicit Chiral Symmetry Breaking}\label{explicite}
When the chiral symmetry gets explicitly broken and a small bare mass of the quark, $m_{_{0}}$, cannot be ignored  in Lagrangian~(\ref{LNJL1}), effect of the rotation $\psi\rightarrow \exp{(i \gamma_5\alpha/2)}\psi $ on such a mass term for small values of $\alpha$ is~\cite{Vogl}
\begin{equation}\label{chiralrot1}
\delta \mathcal{ L}=-m_{_{0}}\bar\psi \left(1+i\gamma_5\alpha-\frac{1}{2}\alpha^2+\cdot\cdot\cdot\right)\psi=-m_{_{0}}\bar\psi \psi+\delta \mathcal{ L}_{\it int}\, .
\end{equation}
Recognizing $\bar\psi  i\gamma_5\psi$ as the source of the pseudo-scalar pion field $\chi$, one can identify $\alpha=\chi/f_{\rm \pi}$, with sub-index small letter ${\rm p}$ standing for pion and $f_{{\rm \pi}}$ being the pion field decay constant, and hence $\alpha^2=\chi^2/f_{{\rm \pi}}^2$. One can see that the second order term has the structure $\,(m_{_{0}}/2f_{{\rm \pi}}^2)\bar\psi\psi\chi^2$ and corresponds to the mass term $\delta\mathcal{L}=-1/2\,m_{{\rm \pi}}^2\chi^2$ provided
\begin{equation}\label{GOR}
m_{{\rm \pi}}^2 f_{{\rm \pi}}^2=-m_{_{0}}\langle\bar\psi \psi\rangle.
\end{equation}
 Therefore, the pion  mass $m_{{\rm \pi}}$ is proportional to $\sqrt{m_{_{0}}}$  and the chiral condensate $\langle\bar\psi\psi\rangle$ that is, the $m_{{\rm \pi}}$ vanishes in the chiral limit.
 This is the Gell-Mann, Oakes, Renner (GOR)~\cite{GOR} relation. It is known empirically that $f_{{\rm \pi}}\approx 93 MeV$ where $f_{{\rm \pi}}$ is related to the chiral symmetry breaking scale~\cite{Scherer1}, $\Lambda_{\chi}\approx1 GeV$, via the relation $\Lambda_{\chi}\approx 4\pi f_{{\rm \pi}}$.
  Taking $m_{_{0}}\approx 7 MeV$ for the quark bare mass~\cite{Beil} and~\cite{Vogl} 
 $\langle\bar\psi \psi\rangle\propto (240\,MeV)^3$ in vacuum, the pion mass becomes
  \begin{equation}\label{pionmass}
  m_{{\rm \pi}} \approx 106 MeV.
 \end{equation}
   This is an approximate value, close to the observed mass $m_{{\rm \pi}}\approx 134.9 MeV.$

\subsection*{Anomaly-induced decay in the NJL model}
An important consequence of the pion becoming massive is that a massive pion can decay to two photons as a manifestation of the triangle anomaly~\cite{Scherer1} via $\pi^{^{0}}\rightarrow 2 \gamma$, where the decay rate is 
\begin{equation}\label{decayrate1}
\Gamma^{\rm Dec}_{\rm \pi}= \frac{\alpha^2_{_{\rm fs}} m_{\pi}^3 n^2}{576 \pi^3 f_{\pi}^2}\approx 7.6\times(\frac{n}{3}) eV,
 \end{equation}
 with $n$ the number of colors and $\alpha_{_{fs}}\approx 1/137$ the fine structure constant. This anomaly-induced pion decay is directly analogous to the pseudo-scalar NG mode decay we predict in Weyl superconductors.

\section*{Appendix B: Covariant form and the Lorentz structure of the 8D BdG Hamiltonian. }
To cast the BdG Hamiltonian~(\ref{BCS/FFLO}) into a covariant form we define the 8 dimensional matrices
\begin{align}\label{8Gamma}
\Gamma_\mu=\rho_1\otimes\gamma_\mu, \quad \Gamma_5 =  \mathbb{I}_{{2}}\otimes\gamma_5,
\quad\textit{and}\quad
\Sigma_i=\rho_i\otimes \mathbb{I}_{{4}},
 \end{align}
 where $\rho_i$ with i=1,2,3 are Pauli matrices in the particle-hole basis and  $\mathbb{I}_{{4}}$ is a $4\times 4$ identity matrix. Also, the matrices
 $\Gamma_{\mu}$ and $\Sigma_{i}$ satisfy the algebra
\begin{align}\label{commut1}
&\{\Gamma_{\mu},\Gamma_{\nu}\}= 2\eta_{\mu\nu}\mathbb{I}_{8},\quad  \{\Sigma{i},\Sigma{j}\}= 2\delta_{ij}\mathbb{I}_{8},
\end{align}
where $\eta_{\mu\nu}$ (with $\mu,\nu= 0,\cdots,3$) is the
Minkowski metric in $(1+3)$ dimensions with the signature $-2$. Using these matrices we can write the BdG Hamiltonian in the covariant form given in~(\ref{covHamil}).

It can be seen that in the limit that the pairing vanishes, the BdG Lagrangian in~(\ref{covHamil}) is invariant under the symmetry transformations
\begin{align}\label{transform11}
&\Psi\to e^{i \Gamma_5 \theta/2} \Psi, \quad
\bar{\Psi} \to \bar{\Psi} e^{i\Gamma_5 \theta/2}
\end{align}
and
\begin{align}\label{transform12}
&\Psi\to e^{i \Sigma_3 \varphi/2} \Psi, \quad
\bar{\Psi} \to \bar{\Psi} e^{i \Sigma_3 \varphi/2},
\end{align}
where $\theta$ and $\varphi$ are arbitrary phases.
To understand the role of the $\Gamma_5$ and $\Sigma_3$ matrices, we act by the projection operators $P_\pm = \frac{1}{2} (\mathbb{I}_8 \pm \Gamma_5), P_{e,h} = \frac{1}{2} (\mathbb{I}_8 \pm \Sigma_3)$, on the spinor~(\ref{nambuspinor}) to get
 \begin{align}
 P_+ P_e \Psi = \Phi_+,\qquad  P_- P_e \Psi = \Phi_-,\qquad  P_+ P_h \Psi = \Phi_+^c,\qquad  P_- P_h \Psi = \Phi_-^c,
 \end{align}
  which suggests that the operators $\Gamma_5$ and $\Sigma_3$ are generators of the $U(1)_A$ axial symmetry and $U(1)$ gauge symmetry leading to the chiral and electric charges, respectively. So, writing the BdG Hamiltonian in an 8-dimensional representation leads to a new generator corresponding to the axial symmetry. 
Using commutation relations 
\begin{align}\label{commut}
[\Sigma_{1},\Gamma_{\mu}]=\mathbf{0}_{8},\qquad \{\Sigma_{2,3},\Gamma_{\mu}\}=\mathbf{0}_{8},\quad [\Sigma_{i},\Gamma_{5}]=\mathbf{0}_{8}\quad\textit{and}\quad\{\Gamma_5,\Gamma_{\mu}\}=\mathbf{0}_{8},
\end{align}
 we observe that $\bigt_{\rm F}$ in Lagrangian~(\ref{covHamil}) is not invariant under neither transformation~(\ref{transform11}) nor~(\ref{transform12}), corresponding to the chiral $U(1)_A$ and gauge $U(1)$ symmetries, respectively and hence breaks both symmetries.

To verify the Lorentz invariance of Hamiltonian and Lagrangian in~(\ref{covHamil}) we define the Lorentz generators in the $8\times8$ representation as
\begin{equation}
J^{\mu\nu}_{(8)} = \tfrac{1}{4}\,[\Gamma^\mu,\Gamma^\nu].
\end{equation}
Using $\Gamma_\mu = \rho_1 \otimes \gamma_\mu$ and $\rho_1^2=\mathbb{I}_2$, we obtain
\begin{equation}
J^{\mu\nu}_{(8)} = \mathbb{I}_2 \otimes J^{\mu\nu}_{(4)}, 
\qquad
J^{\mu\nu}_{(4)} = \tfrac{1}{4}\,[\gamma^\mu,\gamma^\nu].
\end{equation}
Consequently, the Lorentz transformation on the 8-component spinor factorizes as
\begin{equation}
S_{(8)}(\Lambda) 
 = \exp\!\left(\tfrac{1}{2}\omega_{\mu\nu}J^{\mu\nu}_{(8)}\right)
 = \mathbb{I}_2 \otimes S_{(4)}(\Lambda),
\end{equation}
where $S_{(4)}(\Lambda) = \exp\!\left(\tfrac{1}{2}\omega_{\mu\nu}J^{\mu\nu}_{(4)}\right)$ is the usual Dirac spinor transformation in $(1+3)$ dimensions~\cite{peskin}. Hence, the $\Gamma_\mu$ transform covariantly
\begin{equation}
S_{(8)}\,\Gamma^\rho\,S_{(8)}^{-1}
= \rho_1 \otimes \big(S_{(4)}\gamma^\rho S_{(4)}^{-1}\big)
= \Lambda^{\rho}{}_{\sigma}\,\Gamma^\sigma,
\end{equation}
and any $\Sigma_i$ of the form $\Sigma_i = \rho_i \otimes \mathbb{I}_4$ remains invariant
\begin{equation}
S_{(8)}\,\Sigma_i\,S_{(8)}^{-1} = \Sigma_i
\end{equation}
and thus the Hamiltonian~(\ref{covHamil}) remains invariant under the extended transformation $S_{(8)} = \mathbb{I}_2\otimes S_{(4)}$.

\section*{Appendix C: }

Applying Fierz transformations, analogous to Lagrangian~(\ref{LNJL2}) in the NJL model, Lagrangian~(\ref{BdGNJLLAG}) can be extended to include vector and axial-vector channels
\begin{equation}\label{BdGNJLLAGFierz}
{\cal L}_{\rm NJL}^{\rm '~ 3DWS} = \Omega \Big[ -G_V (\bar{\Psi}\Gamma_\mu   \Psi)^2 + G_A (\bar{\Psi} \Gamma_\mu   \Gamma_5 \Psi)^2\Big],
\end{equation}
where $G_V$ and $G_A$ are the corresponding couplings. Both Lagrangians~(\ref{BdGNJLLAG}) and~(\ref{BdGNJLLAGFierz}) are invariant under the symmetry transformations corresponding to the generator $\Gamma_5$ due to the commutation relations~(\ref{commut2}).
The corresponding bound states can be found by analyzing the T-matrix in the scalar and $\Gamma_5$ channel. The pole equation determining the collective mode masses reads
\begin{equation}\label{pole1}
1 - K \, \Pi_Y^{\rm 3DWS}(Q^2 = M_{SM}^2) = 0,
\end{equation}
where the index ${SM}$ denotes the superconductive mesons, $\Pi_Y^{\rm 3DWS}$ is the polarization function for a given quasi-particle channel, and $Y=S,P,V,A$ labels the scalar, pseudo-scalar, vector, and axial-vector channels. Explicitly,
\begin{equation}
\Pi_Y^{\rm 3DWS}(Q^2) = i \int^\Lambda \frac{d^4 P}{(2\pi)^4} \mathrm{Tr}[\Gamma_Y S(P+Q) \Gamma_Y^\dagger S(P)],
\end{equation}
with the quasi-particle propagator
\begin{equation}
S(P) = \frac{\Gamma_\mu P^\mu +   \bigt_{_{\rm F}} \Omega }{P^2 - \bigt_{_{\rm F}} ^2 + i \epsilon},
\end{equation}
and
\begin{equation}
\Gamma_S =   \mathbb{I}_{{8}} , \quad \Gamma_P = \Gamma_5, \quad \Gamma_V = \Gamma_\mu  , \quad \Gamma_A = \Gamma_\mu   \Gamma_5.
\end{equation}
Hence, besides the scalar phase mode and amplitude mode in the scalar channel (well-known in conventional superconductors~\cite{Littlewood}) and the pseudo-scalar NG mode in $\Gamma_P$, one expects a vector mode in $\Gamma_V$, an axial-vector mode in $\Gamma_A$, and a Higgs-like amplitude mode in $\Gamma_P$.


\begin{thebibliography}{99}
 \bibitem{Weinberg}S. Weinberg, ``The Quantum Theory of Fields'', vol. {\bf II}, Cambridge University Press, Cambridge, (1996).
\bibitem{BCS}J. Bardeen, L. N. Cooper, and J. R. Schrieffer, {\it Phys. Rev.} {\bf 108}, 1175 (1957).
\bibitem{Nambu1}Y.~Namu,``Spontaneous Symmetry Breaking in Particle Physics: a Case of Cross Fertilization,'' Nobel Lecture, (2008).
  \bibitem{Nambu2} Y. Nambu, {\it Phys. Rev.} {\bf 117}, 648 (1960). Y. Nambu, {\it Phys. Rev. Lett.} {\bf 4} , 380 (1960).
     \bibitem{Littlewood}P. B. Littlewood and C. M. Varma, {\it Phys. Rev. B.} {\bf 26}, 4883 (1982).
\bibitem{Anderson1} P. W. Anderson, {\it Phys. Rev.} {\bf 110}, 827  (1958).
\bibitem{Anderson2}  P. W. Anderson, {\it Phys. Rev.} {\bf 112}, 1900 (1958).
\bibitem{Higgs} P. W. Higgs, {\it Phys. Lett.} {\bf 12}, 132 (1964).
  \bibitem{Meissner} W. Meissner and R. Ochsenfeld, {\it Naturwissenschaften } {\bf 21}, 787 (1933).
  \bibitem{Bednik2016} G. Bednik, A.A. Zyuzin and A.A. Burkov, {\it New J. Phys.} {\bf 18}, 085002 (2016).
  \bibitem{AFM2025} M. Z-Abyaneh, M. Farhoudi, M. Mashkoori, {\it Sci. Rep. } {\bf 15}, 13803 (2025).
\bibitem{FFLO} P. Fulde and R. A. Ferrell, {\it Phys. Rev.} {\bf 135}, A550 (1964).
\bibitem{Cho2012} G. Y. Cho, J. H. Bardarson, Y. M. Lu, and J. E. Moore,{\it Phys. Rev. B} {\bf 86}, 214514 (2012).
  \bibitem{Bednik2015}  G. Bednik, A. A. Zyuzin, and A. A. Burkov, {\it Phys. Rev. B} {\bf92}, 035153 (2015).
\bibitem{Matsuda2007} Y. Matsuda and H. Shimahara, {\it J. Phys. Soc. Jpn.} {\bf 76}, 051005 (2007).

  \bibitem{NJL1}Y. Nambu and G. Jona-Lasinio, {\it Phys. Rev.} {\bf 122},  345 (1961).
\bibitem{NJL2}Y. Nambu and G. Jona-Lasinio, {\it Phys. Rev.} {\bf124}, 246 (1961).
\bibitem{Valatin} G. Valatin, {\it Nuovo Cimento} {\bf 7}, 843 (1958).
\bibitem{Gennes}  P. G. de Gennes, ``Superconductivity of Metals and Alloys'',  Addison-Wesley Publishing, New York, (1966).
\bibitem{Bogoliubov1} N.N. Bogoliubov, {\it Sov. Phys. JETP} {\bf 7}, 41 (1958).
  \bibitem{dirac2-1928}P. A. M. Dirac, {\it Proc. Roy. Soc.}\ {\bf A117}, 610 (1928).; {\it ibid}\ {\bf A118}, 351 (1928).
  \bibitem{Scherer1} S. Scherer, ``Introduction to Chiral Perturbation Theory," arXiv:hep-ph/0210398v1.
  \bibitem{thesisAbyaneh} M. Z. Abyaneh, Some Aspects of Chiral Perturbation Theory and Neutrino Physics, PhD Thesis.
\bibitem{GOR} M. Gell-Mann, R. Oakes and B. Renner, {\it Phys.Rev.} {\bf 175}, 2195 (1968).
\bibitem{Creti24} C. Corian\'o, M. Cret\'i, S. Lionetti, and R. Tommasi, ``Axion-like Quasiparticles and Topological States of Matter: Finite Density Corrections of the Chiral Anomaly Vertex" (2024). arXiv:2402.03151 
\bibitem{Palumbo25} G. Palumbo, ``Boundary Witten effect in multi-axion insulators" (2025). arXiv:2504.16919 
\bibitem{Boyanovsky25} D. Boyanovsky, {\it Phys. Rev. B} {\bf 112}, 174301 (2025). 
\bibitem{Shyta25} V. Shyta, J. van den Brink, and F. S. Nogueira, {\it Phys. Rev. B} {\bf 112}, 104517 (2025). 
\bibitem{Mottola2024} E. Mottola, A. V. Sadofyev, and A. Stergiou, {\it Phys. Rev. B} {\bf 109}, 134512 (2024).
\bibitem{Bernabeu23} J. Bernabeu and A. Cortijo, {\it arXiv preprint} arXiv:2311.07644 (2023).
\bibitem{Lhachemi23} M. Nabil, M. N. Y. Lhachemi and I. Garate, {\it Phys. Rev. B} {\bf 109}, 144304 (2024). 

  \bibitem{Klein1984} M. V. Klein, S. B. Dierker, {\it Phys. Rev. B} {\bf 29}, 4976 (1984).

\bibitem{Shimino}R. Shimano, N. Tsuji, {\it Annu. Rev. Condens. Matter Phys.} {\bf 11}, 103 (2020).
\bibitem{Liu2016} Q. Liu {\it et al.}, {\it Nat. Commun.} {\bf 7}, 11038 (2016). 
\bibitem{Jiang2017} J. Jiang {\it et al.}, {\it Nat. Commun.} {\bf 8}, 13973 (2017).
\bibitem{Xu2015} S.-Y. Xu {\it et al.}, {\it Sci. Adv.} {\bf 1}, e1501092 (2015).
\bibitem{Sinha2020} D. Sinha, {\it Phys. Rev. B} {\bf 102}, 085144 (2020).
\bibitem{Dutta2020} P. Dutta, F. Parhizgar, and A. M. Black-Schaffer, {\it Phys. Rev. B} {\bf 101}, 064514 (2020).
\bibitem{ZyuzinBurkov2012} A. A. Zyuzin and A. A. Burkov, {\it Phys. Rev. B} {\bf 86}, 115133 (2012).
\bibitem{Burkov2} A. A. Burkov, {\it J. Phys.: Condens. Matter} {\bf 27}, 113201 (2015).
\bibitem{Ding} Y. Ding et.al, {\it Nano Lett.} {\bf 24}, 41 (2024).
\bibitem{Ishikawa1984}K. Ishikawa,{\it Phys. Rev. Lett.} {\bf 53}, 1615 (1984).
\bibitem{Goryo1999}J. Goryo and K. Ishikawa, {\it Phys. Lett. A} {\bf 260}, 294 (1999).
\bibitem{Morimoto2016} T. Morimoto, S. Zhong, J. Orenstein, and J. E. Moore, {\it Phys. Rev. B} {\bf 94}, 245121 (2016).
\bibitem{Vogl} U. Vogl and W. Weise, {\it Prog. Part. Nucl. Phys.} {\bf 27}, 195 (1991).
\bibitem{Beil} C. Beil, {\it A derivation of the first generation particle masses from internal spacetime}, arXiv:2405.15522 (2024).
\bibitem{peskin} M. Peskin and D. Schroeder, {\it An Introduction to Quantum Field Theory}, Westview Press, Chicago (1995).



 \end{thebibliography}
\end{document}